\documentclass[a4paper,twocolumn,11pt,unpublished]{quantumarticle}
\pdfoutput=1
\usepackage[T1]{fontenc}
\usepackage[utf8]{inputenc}
\usepackage{amsmath,amssymb,amsthm}
\usepackage{booktabs,array}
\usepackage[numbers,sort&compress]{natbib}
\usepackage{microtype}
\usepackage{hyperref}
\usepackage{tikz}
\usetikzlibrary{arrows.meta,positioning,calc}
\usepackage{enumitem,needspace,placeins}
\setlist{itemsep=2pt,topsep=4pt,leftmargin=*}
\newtheorem{theorem}{Theorem}
\newtheorem{proposition}[theorem]{Proposition}
\newtheorem{lemma}[theorem]{Lemma}
\theoremstyle{definition}

\newcommand{\F}{\mathbb F_2}
\newcommand{\ps}[1]{\mathsf{#1}}
\newcommand{\GS}{\mathcal G(S)}
\newcommand{\supp}{\operatorname{supp}}
\newcommand{\ecc}{\operatorname{ecc}}
\newcommand{\dist}{\operatorname{dist}}
\newcommand{\rank}{\operatorname{rank}}
\newcommand{\Nbr}{\mathcal N}
\newcommand{\spn}{\operatorname{span}}
\makeatletter
\pretocmd{\@printauthor}{\ifnumcomp{#1}{=}{4}{\newline}{}}{}
  {\PackageError{fusionauthors}{Could not set the author line break}{Check the quantumarticle author macro.}}
\makeatother
\begin{document}
\raggedbottom
\hypersetup{linkcolor=blue!50!black,citecolor=blue!50!black,urlcolor=blue!50!black}
\title{Optimal Measurement Rounds for Perfect Encoded Fusion}
\author{Seita Tsuboi\texorpdfstring{\textsuperscript{*}}{}}
\email{tsuboi@mtl.t.u-tokyo.ac.jp}
\author{Takuya Kasamura\texorpdfstring{\textsuperscript{*}}{}}
\author{Leo Itoh\texorpdfstring{\textsuperscript{*}}{}}
\author{Motoharu Tsuboi\texorpdfstring{\textsuperscript{*}}{}}
\author{Tomohiro Hatori\texorpdfstring{\textsuperscript{*}}{}}
\author{Junichiro Kadomoto\texorpdfstring{\textsuperscript{*}}{}}
\affiliation{\textsuperscript{*}The University of Tokyo}
\date{}
\begin{abstract}
Encoded fusion combines probabilistic physical fusions to perform a logical Bell
measurement. Adaptive choices depend on earlier measurement outcomes and therefore
require multiple measurement rounds. We determine the minimum worst-case number
of rounds for perfect fusion of two blocks of an irreducible $[[n,1]]$ stabilizer
code. A strategy is perfect if the logical Bell measurement succeeds on every
branch containing at least one successful physical fusion. Even after fixing the
past history, including measurement values, we consider all success/failure
branches, including those with zero probability for a particular input state or
measurement device. In a lossless model with fixed coordinate pairing, arbitrary
Pauli failure bases, and each pair fused exactly once, the optimum equals the
minimum root eccentricity over the local-complementation orbit of rooted graphs
representing the code. The lower bound applies to arbitrary perfect strategies
that choose batches and bases using previous outcomes. We also characterize the
batch sequences that can occur on an all-failure branch of a perfect strategy by
a necessary and sufficient graph adjacency condition. A matching strategy fuses
distance layers from the outside in and, after the first success, completes the
remaining fusions in at most one further round by changing the failure bases
along an induced path.

\end{abstract}
\maketitle
\hypersetup{pdfauthor={Seita Tsuboi, Takuya Kasamura, Leo Itoh, Motoharu Tsuboi, Tomohiro Hatori, Junichiro Kadomoto}}
\section{Introduction}\label{sec:introduction}

Fusion-based quantum computation connects small photonic resource states through
measurements~\cite{fbqc}. Physical fusion is probabilistic: success reveals two
Bell parities, whereas failure measures the two photons individually in a chosen
Pauli basis. Encoding can make a logical Bell measurement succeed even when some
physical fusions fail. In adaptive fusion, feed-forward (FF) lets earlier
outcomes determine subsequent batches and failure bases. Acquiring and processing
these outcomes may require delaying unmeasured photons, increasing latency and
exposure to loss. Fewer rounds therefore reduce sequential control requirements.

This leads to a question about measurement rounds. For a fixed code, what is the
minimum worst-case number of rounds needed for logical fusion to succeed on every
branch containing at least one physical success? We call a strategy satisfying
this condition \emph{perfect}.

Let $R^*(S)$ denote the minimum worst-case number of rounds among perfect
strategies for a code $S$. For irreducible $[[n,1]]$ stabilizer codes, we prove
\[
 R^*(S)=\min_{\Gamma\in\GS}\ecc_\Gamma(e).
\]
Here the root $e$ represents the encoded qubit, and $\GS$ is the set of rooted
graph representations of the code, which forms a single local-complementation
orbit. The root eccentricity $\ecc_\Gamma(e)$ is the largest shortest-path
distance from $e$ to a physical vertex. The optimization therefore ranges over
graph representations as well as measurement orders within a fixed graph.

Goodenough \emph{et al.} established recoverability criteria and characterized
perfect nonadaptive fusion~\cite{goodenough}. Rei{\ss} and van Loock gave adaptive
strategies, including layer constructions for tree codes~\cite{reiss}; a
sequential construction for connected graph codes appears in a machine-generated
Lean proof made public before this work~\cite{pr13260}. We characterize the
minimum worst-case round count over arbitrary perfect adaptive strategies for
irreducible $[[n,1]]$ codes in the full outcome-tree model, proving a matching
lower bound and graph-based construction, together with a necessary and sufficient
condition on all-failure batch sequences.

Section~\ref{sec:model} defines the model and states the main theorem.
Sections~\ref{sec:criterion} and~\ref{sec:lower} develop the recovery criterion
and necessary conditions for perfect strategies.
Section~\ref{sec:graphs} provides the graph representations and induced-path
completion used in Section~\ref{sec:schedules} to prove the batch
characterization and the main theorem. Section~\ref{sec:example} illustrates
optimization over graph representations and a code that requires two rounds.
Section~\ref{sec:conclusion} concludes.

\section{Model and Main Result}\label{sec:model}

We define the code and fusion strategy model, then introduce rooted graph
representations and state the optimal-round theorem with its proof roadmap.

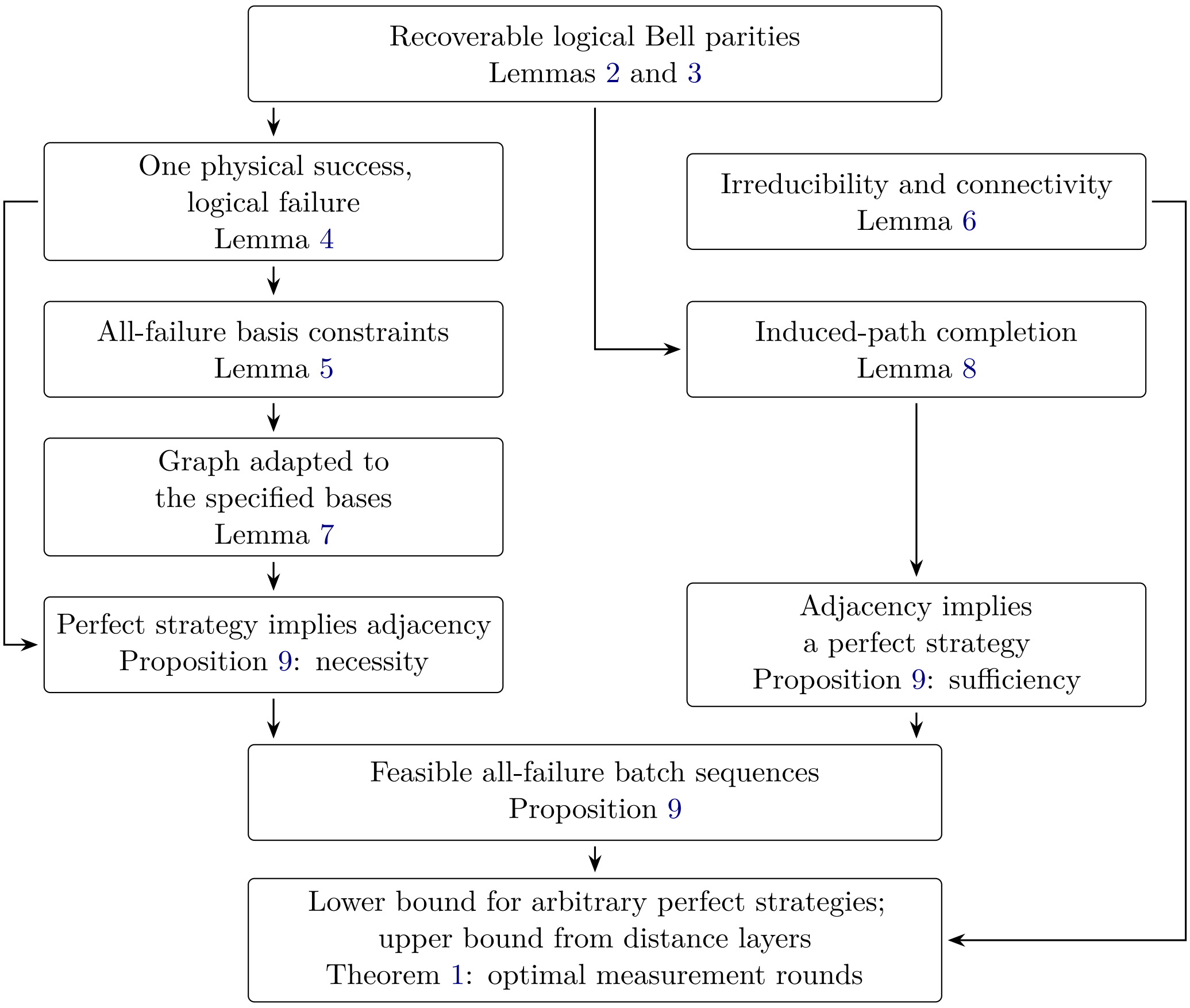
\begin{figure*}[t]
\centering
\begin{tikzpicture}[
 box/.style={draw,rounded corners=2pt,align=center,text width=5.0cm,
   minimum height=1.1cm,inner sep=4pt,font=\fontsize{9.5pt}{12pt}\selectfont},
 wide/.style={box,text width=7.7cm},
 arr/.style={-{Stealth[length=2mm]},line width=0.6pt},
 every node/.style={outer sep=2pt}]
 \node[wide] (recovery) at (0,0)
   {Recoverable logical Bell parities\\Lemmas~\ref{lem:criterion} and~\ref{lem:full}};
 \node[box] (isolated) at (-3.7,-1.7)
   {One physical success,\\logical failure\\Lemma~\ref{lem:isolated}};
 \node[box] (connected) at (3.7,-1.7)
   {Irreducibility and connectivity\\Lemma~\ref{lem:connected}};
 \node[box] (constraint) at (-3.7,-3.4)
   {All-failure basis constraints\\Lemma~\ref{lem:transverse}};
 \node[box] (path) at (3.7,-3.4)
   {Induced-path completion\\Lemma~\ref{lem:path}};
 \node[box] (adapted) at (-3.7,-5.1)
   {Graph adapted to the specified bases\\Lemma~\ref{lem:graphform}};
 \node[box] (sufficient) at (3.7,-6.8)
   {Adjacency implies a perfect strategy\\Proposition~\ref{prop:schedules}: sufficiency};
 \node[box] (necessary) at (-3.7,-6.8)
   {Perfect strategy implies adjacency\\Proposition~\ref{prop:schedules}: necessity};
 \node[wide] (batches) at (0,-8.5)
   {Feasible all-failure batch sequences\\Proposition~\ref{prop:schedules}};
 \node[wide] (main) at (0,-10.2)
   {Lower bound for arbitrary perfect strategies;\\upper bound from distance layers\\Theorem~\ref{thm:main}: optimal measurement rounds};
 \draw[arr] (recovery.south -| isolated.north)--(isolated.north);
 \draw[arr] (isolated)--(constraint);
 \draw[arr] (constraint)--(adapted);
 \draw[arr] (adapted)--(necessary);
 \draw[arr] (recovery.south) -- (0,-3.4) -- (path.west);
 \draw[arr] (path)--(sufficient);
 \draw[arr] (isolated.west) -- (-6.8,-1.7) -- (-6.8,-6.8) -- (necessary.west);
 \draw[arr] (necessary.south)--(batches.north -| necessary.south);
 \draw[arr] (sufficient.south)--(batches.north -| sufficient.south);
 \draw[arr] (batches)--(main);
 \draw[arr] (connected.east) -- (6.8,-1.7) -- (6.8,-10.2) -- (main.east);
\end{tikzpicture}
\caption{Main dependencies in the proof. The left side gives necessary conditions
for arbitrary perfect strategies; the right side constructs strategies.
Connectivity is used for the distance-layer upper bound and for nonexistence
for reducible codes.}
\label{fig:proof_dependencies}
\end{figure*}

\subsection{Stabilizer Codes and Adaptive Fusion}\label{sec:roundmodel}

Write $\F$ for the binary field; all vector and matrix arithmetic below is over
this field. Pauli words on $n$ physical qubits, modulo phase, are represented by
vectors in $V=(\F^2)^n$. We use the equivalent notations
\[
 v=((x_1,z_1),\ldots,(x_n,z_n))\equiv(x\mid z),
\]
where $x,z\in\F^n$ and $v_i=(x_i,z_i)\in\F^2$ is the two-bit component at
coordinate $i$. At one coordinate,
\[
 \begin{aligned}
 \ps X&=(1,0),&\ps Z&=(0,1),\\
 \ps Y&=(1,1),&\ps I&=(0,0).
 \end{aligned}
\]
Vector addition corresponds to multiplication of Pauli words modulo phase.
We write $[n]=\{1,2,\ldots,n\}$.
The \emph{support} is
\[
 \supp(v)=\{i\in[n]:v_i\ne(0,0)\}.
\]
For $A\subseteq[n]$, let $V_A$ be the subspace of vectors that vanish outside $A$.

The symplectic form
\[
 \omega((x,z),(x',z'))=x\cdot z'+z\cdot x'
\]
is zero for commuting Pauli words and one for anticommuting words. An $[[n,1]]$
stabilizer code, encoding one logical qubit, is represented by an isotropic
subspace $S\subset V$ of dimension $n-1$. Isotropy means
$\omega(s,t)=0$ for all $s,t\in S$, expressing commutativity of the stabilizers.

Let $N=S^{\perp_V}$ be the space of Pauli words commuting with every stabilizer.
Physical Pauli words that differ by a stabilizer represent the same logical
operator. Their classes are
\[
 [u]=u+S\in N/S\qquad(u\in N).
\]
The quotient $N/S$ is a two-dimensional nondegenerate symplectic space with form
\[
 \bar\omega([u],[v])=\omega(u,v).
\]
We call its elements \emph{logical Pauli classes}; nonzero classes represent
nontrivial logical Pauli operators. A physical representative $u\in N\setminus S$
is a \emph{nontrivial logical word}, or simply a logical word. Thus $u$ is a
vector in the physical space $V$, whereas $[u]$ belongs to the quotient.

We call $S$ \emph{irreducible} if
\begin{equation}\label{eq:irreducible}
 \dim(S\cap V_A)<|A|
 \quad(\varnothing\ne A\subseteq[n]).
\end{equation}
This excludes a detached subsystem in a pure stabilizer state that is independent
of the logical information.

We fuse two blocks of the same code with fixed coordinate pairing. Coordinate
$i$ denotes a pair containing one photon from each block. Its \emph{failure
basis} is chosen from $\ps X,\ps Y,\ps Z$. A successful physical fusion reveals
two Bell parities; on failure, the photons are measured individually in the
failure basis. Photon loss is excluded.

In each round a fusion strategy chooses a nonempty \emph{batch} of previously
unfused coordinates and a failure basis for each. A batch is the set of
coordinates fused in the same round. These choices are fixed before any outcome
of that round is known. All earlier outcomes, including individual measurement
values, may be used. We call such outcome-dependent choices \emph{adaptive
fusion}, and the use of outcomes from earlier rounds to select subsequent
measurements \emph{feed-forward (FF)}.

Every coordinate is fused exactly once on every branch; a strategy does not stop
early even if the logical information is already recoverable. The pairing is
fixed. Ancillary photons, single-photon measurements used to complete fusion,
and operations that boost the physical fusion success probability are not
allowed.

We now specify the outcome branches on which a fusion strategy is required to succeed.
We use the \emph{full outcome-tree model}: after fixing the past history,
including measurement values, every success/failure assignment in the next
batch is included as a branch. A \emph{perfect fusion strategy}, or perfect
strategy, recovers both logical Bell parities on every such branch containing
at least one successful physical fusion. Branches with zero probability for a
particular input state and measurement device are not excluded. Measurement
values may affect control choices, but not the recovery criterion in
Section~\ref{sec:criterion}. Phase-free vectors specify which parities are
recoverable; their measured signs are tracked separately by classical processing.

Let $\Pi_{\mathrm{perf}}(S)$ be the set of perfect fusion strategies for $S$, and
let $\mathcal B(\pi)$ be the set of outcome branches of a strategy $\pi$.
Write $R_\pi(b)$ for the number of nonempty batches executed on branch
$b\in\mathcal B(\pi)$. We minimize the worst-case number of rounds:
\begin{equation}\label{eq:Rstar}
 R^*(S)=\min_{\pi\in\Pi_{\mathrm{perf}}(S)}\;\max_{b\in\mathcal B(\pi)}R_\pi(b).
\end{equation}
Set $R^*(S)=\infty$ if no perfect strategy exists. One round requires no FF;
$R$ rounds use at most $R-1$ inter-round FF steps. It suffices to consider
deterministic strategies: fixing the internal random choices of a strategy
that is perfect on every branch preserves perfection and cannot increase the
worst-case number of rounds.

\subsection{Rooted Graphs and Optimal Measurement Rounds}

Graph representations allow rounds to be constructed from distances to a root.
Let $\Gamma$ be a simple graph on the fixed labeled vertex set
$\{e\}\sqcup[n]$, with $e$ nonisolated. Use the same symbol for its binary
adjacency matrix. The graph-state stabilizer subspace is
\[
 W_\Gamma=\{(a,\Gamma a):a\in\F^{n+1}\},
\]
generated by $h_v=\ps X_v\prod_{u\in\Nbr_\Gamma(v)}\ps Z_u$, where
$\Nbr_\Gamma(v)$ is the neighborhood of $v$. The corresponding graph code is
\begin{equation}\label{eq:graphcode}
 S_\Gamma=\{w|_{[n]}:w\in W_\Gamma,\ w_e=0\}.
\end{equation}
The root $e$ represents the encoded qubit and is not a physical fusion
coordinate. Equation~\eqref{eq:graphcode} selects the graph-state words that
are trivial at the root and restricts them to the physical coordinates.

Coordinatewise symplectic maps independently permute $\ps X,\ps Y,\ps Z$ at
each physical coordinate. Let $\GS$ be the set of rooted graphs whose graph
codes are related to $S$ by such maps. This set is nonempty and is exactly one
local-complementation orbit, as explained in Section~\ref{sec:graphfacts}.
\emph{Local complementation} at a vertex $v$ toggles the edges between distinct
neighbors of $v$. Local complementation at the root is allowed, but vertex
permutations are not included in the operations generating this orbit. The
\emph{root eccentricity} is
\[
 \ecc_\Gamma(e)=\max_{i\in[n]}\dist_\Gamma(e,i),
\]
the largest shortest-path distance from the root to a physical vertex.

\subsection{Main Theorem and Proof Roadmap}

\begin{theorem}[Optimal measurement rounds]\label{thm:main}
In the full outcome-tree model of Section~\ref{sec:roundmodel},
every irreducible $[[n,1]]$ code with $n\ge1$ satisfies
\begin{equation}\label{eq:main}
 \boxed{R^*(S)=\min_{\Gamma\in\GS}\ecc_\Gamma(e).}
\end{equation}
Reducible codes admit no perfect strategy in this model.
\end{theorem}

In particular, the one-round case is
\[
 R^*(S)=1\quad\Longleftrightarrow\quad
 \exists\Gamma\in\GS:\ \Nbr_\Gamma(e)=[n].
\]
This recovers the characterization of perfect nonadaptive fusion in
Ref.~\cite[Corollary~3.5 and Theorem~4.3]{goodenough}.

The theorem optimizes both the graph representation of the code and the
measurement order. Its proof relies on the characterization of all-failure
batch sequences in Proposition~\ref{prop:schedules}.

\paragraph{Proof roadmap.}
Lemmas~\ref{lem:criterion} and~\ref{lem:full} express recoverable logical Bell
parities in linear-algebraic terms. From this common starting point, we separate
the lower bound from the construction attaining it.

\emph{Lower bound.}
The obstruction to logical success with a single physical success
(Lemma~\ref{lem:isolated}) implies $S\cap F_q=\{0\}$ on every all-failure
branch of a perfect strategy (Lemma~\ref{lem:transverse}). A graph representation
adapted to these specified bases (Lemma~\ref{lem:graphform}) then forces each
vertex to be adjacent either to the root or to a vertex in a later batch. This
is the necessity direction of Proposition~\ref{prop:schedules} and yields a
lower bound for arbitrary perfect strategies.

\emph{Upper bound.}
The adjacency condition provides an induced path from a successful coordinate
in the first round containing a success to the root, with only unfused internal
vertices. Lemma~\ref{lem:path} completes
the remaining fusions in at most one further round, proving sufficiency in
Proposition~\ref{prop:schedules}. Irreducibility is equivalent to connectivity
(Lemma~\ref{lem:connected}); fusing distance layers from the outside in then
gives the main theorem's upper bound. Figure~\ref{fig:proof_dependencies}
summarizes these dependencies.

\section{Logical Fusion and Recoverability}\label{sec:criterion}

We determine which logical Bell parities can be recovered from the physical
fusion outcomes on a branch. We first represent these outcomes by a space of
Pauli words, then characterize logical fusion success by whether the
failure-basis span contains a logical word.

\subsection{Recoverable Logical Bell Parities}

Let $D\subseteq[n]$ be the failed coordinates on a branch, and write
$r=(r_i)_{i\in D}$ for their chosen failure bases. At each $i\in D$, the Pauli
$r_i$ is measured individually in each block. Represent these operators by
phase-free vectors. We call the binary linear subspace they generate,
\begin{equation}\label{eq:frame}
  F_r=\left\{\sum_{i\in D}c_i\iota_i(r_i):c_i\in\F\right\},
\end{equation}
the \emph{failure-basis span}. Here $\iota_i$ embeds a one-coordinate vector
into coordinate $i$ of $V$:
\[
  \iota_i(p)=(0,\ldots,0,p,0,\ldots,0).
\]
Each coefficient $c_i$ specifies whether the Pauli $r_i$ occurs in the product.
Thus a word in $F_r$ has letter $0$ or $r_i$ at each $i\in D$ and vanishes
outside $D$. This space includes products of the directly measured Pauli words.
The definition does not add words inferred using known stabilizer values.

For $u\in F_r$, the value of the corresponding Pauli word is known separately
in each block. Multiplying these two values gives the inter-block parity for
that word. A successful fusion instead gives two independent Bell parities,
without giving the individual Pauli values in each block.
Let $J=[n]\setminus D$ be the successful coordinates.
The physical measurement outcomes determine the inter-block parity associated
with every Pauli word in
\[
  M=V_J+F_r.
\]
For words in $F_r$, the Pauli measurement outcomes are also known separately
for each block.

We now pass from physical words to logical Pauli classes. We retain the words
in $M$ that belong to $N$ and identify words that differ by a stabilizer.
The recoverable logical Bell parities are described by
\begin{equation}\label{eq:BQ}
  \begin{aligned}
    B&=(N\cap M)+S\subseteq N,\\
    Q&=B/S\subseteq N/S.
  \end{aligned}
\end{equation}
Here $B$ is a space of physical representatives, whereas $Q$ is a space of
logical Pauli classes; they belong to different ambient spaces.
The measured signs $\pm1$ give the
parity values, while the subspace $Q$ specifies which parities are recoverable.
Logical Bell measurement succeeds exactly when $Q=N/S$.
Two independent logical Pauli classes need not commute on one block, but their
corresponding inter-block parity operators do commute.

\subsection{Conditions for Logical Fusion Success}

To decide success, it suffices to check whether the failure-basis span
$F_r\subseteq V$ contains a logical word, that is, a physical representative
in $N\setminus S$. The following criterion is Proposition~3.1 of
Goodenough \emph{et al.}~\cite{goodenough}, expressed in the present notation.

\begin{lemma}[Failure-basis span criterion]\label{lem:criterion}
Logical fusion succeeds on a branch if and only if
\begin{equation}\label{eq:complete}
  F_r\cap N\subseteq S.
\end{equation}
Equivalently, $F_r$ contains no element of $N\setminus S$.
\end{lemma}

This condition means that products of the Paulis directly measured at failed
coordinates cannot form a nontrivial logical word.

\begin{proof}
Rather than test $Q=N/S$ directly, we compute the orthogonal complement of $Q$
in the logical quotient. Every orthogonal complement with subscript $V$ below
is taken in the physical space $V$.
At a single coordinate, the symplectic orthogonal complement of $\F^2$ is
$\{0\}$, corresponding to the identity Pauli $\ps I$, while that of any
one-dimensional subspace is the subspace itself. The local subspace of $M$ is
$\F^2$ at each successful coordinate and $\F r_i$ at each failed coordinate,
so
\[
  M^{\perp_V}=F_r.
\]
The identities for subspaces $A,C\subseteq V$,
\[
  \begin{aligned}
    (A+C)^{\perp_V}&=A^{\perp_V}\cap C^{\perp_V},\\
    (A\cap C)^{\perp_V}&=A^{\perp_V}+C^{\perp_V},
  \end{aligned}
\]
together with $N^{\perp_V}=S$, give
\begin{align}
  B^{\perp_V}
   &=\bigl((N\cap M)+S\bigr)^{\perp_V}\notag\\
   &=(S+F_r)\cap N\notag\\
   &=S+(F_r\cap N).\label{eq:Bperp}
\end{align}
The last equality uses $S\subseteq N$.
Moreover, $S\subseteq B\subseteq N$ implies
$S\subseteq B^{\perp_V}\subseteq N$. Hence the orthogonal complement in the
logical quotient $N/S$ is
\[
  Q^{\perp_{\bar\omega}}=B^{\perp_V}/S.
\]
Since the form on $N/S$ is nondegenerate,
\[
  \begin{aligned}
    Q=N/S
    &\quad\Longleftrightarrow\quad Q^{\perp_{\bar\omega}}=0\\
    &\quad\Longleftrightarrow\quad B^{\perp_V}=S.
  \end{aligned}
\]
By Eq.~\eqref{eq:Bperp}, the last condition is equivalent to
$F_r\cap N\subseteq S$.
\end{proof}

Next, independently of the actual success or failure outcomes, consider an
assignment $q=(q_i)_{i\in[n]}$ of one Pauli basis to every coordinate.
The associated failure-basis span $F_q$ is Eq.~\eqref{eq:frame} with $D=[n]$.
It is a maximal isotropic, or Lagrangian, subspace satisfying
\[
  \dim F_q=n,\qquad F_q^{\perp_V}=F_q.
\]
Here $q$ specifies bases; it does not assert that all physical fusions have
actually failed. If an all-failure branch uses these bases, its failure-basis
span is $F_q$.

The next lemma is the $k=1$ case of Lemma~3.2 of Goodenough \emph{et al.}~\cite{goodenough}. We include its short proof to keep the argument self-contained.

\begin{lemma}[A logical word for every full basis assignment]\label{lem:full}
For every assignment $q$ of bases to all coordinates,
\begin{equation}\label{eq:dimension}
  \dim(F_q\cap N)=1+\dim(F_q\cap S).
\end{equation}
In particular, $F_q\cap(N\setminus S)\ne\varnothing$.
\end{lemma}

Thus, for any full basis assignment, the products of the assigned Paulis
include a nontrivial logical word.

\begin{proof}
Since $F_q$ is a Lagrangian subspace of dimension $n$,
\[
  (F_q+S)^{\perp_V}=F_q\cap N.
\]
Applying the dimension formulas for a sum and an orthogonal complement gives
\[
  \begin{aligned}
    \dim(F_q+S)&=2n-1-\dim(F_q\cap S),\\
    \dim(F_q\cap N)&=2n-\dim(F_q+S)\\
      &=1+\dim(F_q\cap S).
  \end{aligned}
\]
If $F_q\cap N$ were contained in $S$, then $F_q\cap N=F_q\cap S$,
contradicting the difference of one in their dimensions.
\end{proof}

On an all-failure branch, a logical word in $F_q$ violates the success condition
of Lemma~\ref{lem:criterion}, so logical fusion necessarily fails.
A perfect strategy therefore succeeds on precisely the patterns containing
at least one successful physical fusion. If the physical fusion flags are
independent, with success and failure each having probability $1/2$, the
success probability is $1-2^{-n}$. The round-count theorem itself does not
use a probability distribution.

\section{Necessary Conditions for Perfect Fusion Strategies}\label{sec:lower}

We derive necessary conditions for a perfect strategy by considering branches
with exactly one successful physical fusion. We first give a condition under
which that success does not suffice for logical fusion. We then constrain the
bases on an all-failure branch so that they cannot produce this obstruction.

\subsection{Logical Fusion Failure Despite a Single Successful Physical Fusion}

A single successful physical fusion need not recover both logical Bell
parities. The next lemma identifies such a failure from the form of a
stabilizer.

\begin{lemma}[A condition for logical failure despite one physical success]\label{lem:isolated}
Suppose only coordinate $i$ succeeds. Let $r$ be the chosen failure bases on
$[n]\setminus\{i\}$ and $F_r$ the corresponding failure-basis span.
If some $s\in S$ satisfies
\[
  s_i\ne0,\qquad s-\iota_i(s_i)\in F_r,
\]
then logical fusion fails.
\end{lemma}

A stabilizer that is nonzero at the successful coordinate and aligned with
the failure bases elsewhere lets us construct a logical word in $F_r$
supported away from the successful coordinate.

\begin{proof}
Extend $r$ to an assignment $r'$ on all coordinates by choosing basis $s_i$
at coordinate $i$. Lemma~\ref{lem:full} gives
$w\in F_{r'}\cap(N\setminus S)$.
If $w_i=0$, then $w\in F_r$ already. Otherwise $w_i=s_i$, so adding the
stabilizer $s$ gives
\[
  \begin{aligned}
    (w+s)_i&=s_i+s_i=0,\\
    [w+s]&=[w]\ne0.
  \end{aligned}
\]
The other components remain aligned with the bases $r$, hence
$w+s\in F_r\cap(N\setminus S)$.
In either case, $F_r$ contains a nontrivial logical word, and
Lemma~\ref{lem:criterion} implies logical failure.
\end{proof}

\subsection{Constraints on All-Failure Branches}

We now derive a necessary condition for a perfect fusion strategy by examining
its all-failure branches. Choose such a branch and let $q$ denote the bases
used on it. If $F_q$ contains a nonzero stabilizer, we show that an alternative
outcome branch satisfies the conditions of Lemma~\ref{lem:isolated},
contradicting perfection.

\Needspace{8\baselineskip}
\begin{lemma}[A necessary condition for a perfect strategy]\label{lem:transverse}
Let $q$ be the basis assignment on any all-failure branch of a perfect strategy.
The corresponding failure-basis span $F_q$ satisfies
\begin{equation}\label{eq:transverse}
  S\cap F_q=\{0\}.
\end{equation}
\end{lemma}

Thus products of the Paulis directly measured on an all-failure branch of a
perfect strategy cannot form a nonzero stabilizer.

\begin{proof}
Suppose $0\ne s\in S\cap F_q$. Along this all-failure branch, let $T_k$ be the
\emph{last} batch meeting $\supp s$, and choose $i\in T_k\cap\supp s$.
Follow the same preceding history, including its measured values, make $i$
the only success in $T_k$, and let every subsequent fusion fail.
The full outcome-tree model of Section~\ref{sec:roundmodel} includes this
continuation in the perfection requirement: every success/failure continuation
of a fixed history is included, even if its probability is zero.

Every coordinate in $\supp s\setminus\{i\}$ has already been fused by the end
of $T_k$. Its failure basis is the same as on the original all-failure branch:
the preceding history is unchanged, and all bases in $T_k$ were fixed before
the outcomes of that round were known.
Thus, writing $r$ for the final failure-basis assignment on the new branch,
\[
  s-\iota_i(s_i)\in F_r.
\]
By Lemma~\ref{lem:isolated}, the new branch fails logically despite the success
at coordinate $i$. This contradicts perfection.
\end{proof}

Combining Lemmas~\ref{lem:full} and~\ref{lem:transverse}, we obtain
$\dim(F_q\cap N)=1$ for every all-failure branch of a perfect strategy.
Thus, its failure-basis span contains exactly one nonzero logical Pauli word.

Choosing the last batch meeting $\supp s$ ensures that any later changes of
basis cannot affect coordinates in the support of $s$.
The bases chosen after the success need not agree with those on the original
all-failure branch. If the controller also uses individual measured values,
the same argument applies after fixing those values in the rounds preceding
the modified batch.

\section{Graph Representations and Fusion Strategies}\label{sec:graphs}

This section translates the necessary condition for a perfect fusion strategy
derived in Section~\ref{sec:lower} into a condition on graph adjacency.
We first describe graph representations of a code and then construct a
representation adapted to specified failure bases.
Finally, an induced path from a successful coordinate to the root provides a
way to complete the remaining fusions.

\subsection{Graph Codes and Connectivity}\label{sec:graphfacts}

\paragraph{Graph codes and the reference qubit.}
Let $\pi_e:W_\Gamma\to V_e$ extract the Pauli component at the root.
By Eq.~\eqref{eq:graphcode}, $S_\Gamma$ is obtained by restricting
$\ker\pi_e$, the graph-state stabilizers acting trivially at the root, to the
physical coordinates. Since every element of $\ker\pi_e$ has zero root
component, this restriction is injective and preserves isotropy.
In particular, $\dim S_\Gamma=\dim\ker\pi_e$.

If the root is not isolated, $\pi_e$ is surjective: $h_e$ has root component
$\ps X$, while $h_v$ for any neighbor $v$ has root component $\ps Z$.
Since $\dim W_\Gamma=n+1$ and $\dim V_e=2$, the rank-nullity theorem gives
\[
 \dim S_\Gamma=\dim\ker\pi_e=(n+1)-2=n-1.
\]

To obtain such a representation for a general code $S$, introduce a reference
qubit $e$.
Choose physical representatives $L_X,L_Z\in N$ of logical Pauli classes with
$\omega(L_X,L_Z)=1$, and define, in $\widetilde V:=V_e\oplus V$,
\begin{equation}\label{eq:purification}
 W=(0\oplus S)+\spn\{(\ps X,L_X),(\ps Z,L_Z)\}.
\end{equation}
The space $W$ is the phase-free stabilizer subspace of the pure state obtained
by placing the logical qubit and the reference qubit in a Bell pair.
The root is a reference system used in the analysis; this construction does
not add an ancillary photon to the fusion protocol.
The two added generators commute because their anticommutation at the root
and on the physical coordinates cancels:
\[
 \omega_{\widetilde V}((\ps X,L_X),(\ps Z,L_Z))=1+1=0.
\]
They also commute with $0\oplus S$ because $L_X,L_Z\in N$.
Thus
\[
 \begin{aligned}
 \dim W&=n+1,& W&=W^{\perp_{\widetilde V}},\\
 W\cap(0\oplus V)&=0\oplus S.
 \end{aligned}
\]
We call this construction a purification of $S$.
By the graph-state normal-form theorem, coordinatewise symplectic maps take
$W$ to a graph-state stabilizer subspace~\cite{vandennest}.
The root projection retains rank two, so the resulting root is not isolated.
Taking the root-zero part gives a graph representation of $S$.

\paragraph{Graph representations and the local-complementation orbit.}
The root projection of a purification $W$ has kernel $0\oplus S$ and image
$V_e$.
Each root component $p\in V_e$ therefore determines a unique logical Pauli
class $[L]\in N/S$ of the physical component.
Indeed,
\[
 \begin{aligned}
 (p,L_1),(p,L_2)\in W
 &\implies(0,L_1+L_2)\in0\oplus S\\
 &\implies[L_1]=[L_2].
 \end{aligned}
\]
Identifying the quotient naturally with a subspace of $V_e\oplus(N/S)$ gives
\[
 \begin{aligned}
 W/(0\oplus S)&=\{(p,\phi(p)):p\in V_e\},\\
 \phi&:V_e\xrightarrow{\ \cong\ }N/S.
 \end{aligned}
\]
Here $\cong$ denotes a symplectic isomorphism: $\phi$ associates the root
Pauli operator $p$ with the code's logical Pauli class $\phi(p)$.
Indeed, isotropy of $W$ gives
$\bar\omega(\phi(p),\phi(p'))=\omega_e(p,p')$.
If $\phi_1,\phi_2$ are the isomorphisms for two purifications, the root map
$\phi_2^{-1}\phi_1$ takes the first purification to the second.
Consequently, two graph codes are related by coordinatewise maps on the
physical coordinates if and only if their graph states are related by
coordinatewise maps including the root.
For the reverse implication, take the root-zero part.
By the local-Clifford equivalence theorem for graph states~\cite{vandennest},
this equivalence is generated by sequences of local complementations.
This is why the set $\GS$ defined in Section~\ref{sec:model} forms a single
local-complementation orbit.
A map acting only on the root changes the logical basis but leaves the code
on the physical coordinates unchanged.

\Needspace{7\baselineskip}
\begin{lemma}[Irreducibility and connectivity of graph codes]\label{lem:connected}
The graph code $S_\Gamma$ is irreducible if and only if $\Gamma$ is connected.
\end{lemma}
Thus graph connectivity characterizes the absence of a pure stabilizer
subsystem detached from the logical qubit.

\begin{proof}
Let $\varnothing\ne A\subseteq[n]$, and let $A^c$ denote its complement in
the full vertex set $\{e\}\sqcup[n]$, including the root.
Here $a_A$ denotes the restriction of $a$ to $A$, and $\Gamma_{A^c,A}$ is the
submatrix of the adjacency matrix $\Gamma$ with rows indexed by $A^c$ and
columns indexed by $A$, describing the edges between these two sets.

Every element of $W_\Gamma$ has the form $(a,\Gamma a)$.
For its support to lie entirely within $A$, both its $X$ and $Z$ components
must vanish outside $A$. Thus $a_{A^c}=0$ and $(\Gamma a)_{A^c}=0$.
Since $a_{A^c}=0$, the latter condition reduces to
$(\Gamma a)_{A^c}=\Gamma_{A^c,A}a_A=0$. Hence,
\[
 \supp(a,\Gamma a)\subseteq A
 \iff
 \begin{cases}
 a_{A^c}=0,\\
 \Gamma_{A^c,A}a_A=0.
 \end{cases}
\]
Since $A$ does not contain the root, every such element has zero root component.
Its restriction to the physical coordinates therefore lies in
$S_\Gamma\cap V_A$. Conversely, every element of $S_\Gamma\cap V_A$ uniquely
lifts to an element of $W_\Gamma$ satisfying these conditions by restoring
the zero root component.

Thus $S_\Gamma\cap V_A$ is in one-to-one correspondence with
$\ker\Gamma_{A^c,A}$, giving
\begin{equation}\label{eq:cutrank}
 \dim(S_\Gamma\cap V_A)=|A|-\rank_{\F}\Gamma_{A^c,A}.
\end{equation}
Since $\Gamma_{A^c,A}=0$ exactly when no edge joins $A$ to $A^c$,
\[
 \begin{gathered}
 \dim(S_\Gamma\cap V_A)=|A|\\
 \iff\text{no edge joins $A$ to $A^c$}.
 \end{gathered}
\]
If $\Gamma$ is disconnected, choose a nonempty connected component not
containing the root as $A$. Then $\dim(S_\Gamma\cap V_A)=|A|$, violating
irreducibility. Conversely, if $\Gamma$ is connected, every nonempty
$A\subseteq[n]$ has an edge to $A^c$, so $\dim(S_\Gamma\cap V_A)<|A|$.
This is precisely the irreducibility condition.
\end{proof}

Coordinatewise invertible maps preserve supports and $\dim(S\cap V_A)$.
Every graph representation of an irreducible code is therefore connected.

\subsection{Strategy-Adapted Graph Representations}\label{sec:adaptedgraph}

To prove a lower bound for an arbitrary perfect fusion strategy, putting the
code alone into graph form is not enough.
We must also align the bases chosen along its all-failure branch so that the
branch can be interpreted on the graph.
The next lemma uses the condition from Lemma~\ref{lem:transverse} to construct
a graph representation in which all specified failure bases become $\ps Z$.

For a pure stabilizer state, a full Pauli assignment whose span intersects the
stabilizer subspace only at zero admits a graph representation mapping every
assigned Pauli to $\ps Z$~\cite[Definition~2.1 and Theorem~2.11]{goodenough_lc}.
The proof below also supplies the root-basis choice needed to apply this
correspondence to a code purification.

\Needspace{8\baselineskip}
\begin{lemma}[Graph form adapted to specified failure bases]\label{lem:graphform}
Suppose that the full failure-basis assignment $q=(q_i)_{i\in[n]}$ satisfies
$S\cap F_q=\{0\}$.
Then there exist $\Gamma\in\GS$ and a coordinatewise symplectic map that takes
$S$ to $S_\Gamma$ and simultaneously takes every $q_i$ to $\ps Z$.
\end{lemma}
The all-failure branch of any perfect fusion strategy can thus be represented
as a graph-code strategy whose failure bases are all $\ps Z$.

\begin{proof}
For $W\subseteq\widetilde V$ in Eq.~\eqref{eq:purification}, set
$U=V_e\oplus F_q$.
In $U$, the physical failure bases are fixed by $q$, while the root component
is unrestricted.
Since $F_q^{\perp_V}=F_q$ and $S\cap F_q=\{0\}$,
\[
 \begin{aligned}
 U^{\perp_{\widetilde V}}&=0\oplus F_q,\\
 W\cap U^{\perp_{\widetilde V}}
   &=(0\oplus S)\cap(0\oplus F_q)=\{0\},\\
 (W+U)^{\perp_{\widetilde V}}
   &=W\cap U^{\perp_{\widetilde V}}=\{0\}.
 \end{aligned}
\]
The last line uses $W^{\perp_{\widetilde V}}=W$.
Hence $W+U=\widetilde V$, and
\[
 \begin{aligned}
 \dim(W\cap U)
 &=\dim W+\dim U-\dim\widetilde V\\
 &=(n+1)+(n+2)-(2n+2)=1.
 \end{aligned}
\]
Write this one-dimensional space as $W\cap U=\F(p,L)$.
The root component $p$ of its nonzero element is nonzero: otherwise
$L\in S\cap F_q=\{0\}$.
Choose
\[
 q_e\in\F^2\setminus\{0,p\},\qquad
 \widetilde F=\F q_e\oplus F_q\subseteq U.
\]
This choice will allow us to obtain a graph-state representation while keeping
all specified bases mapped to $\ps Z$.
\[
 W\cap\widetilde F\subseteq W\cap U=\{(0,0),(p,L)\}.
\]
Since $(p,L)\notin\widetilde F$, we obtain $W\cap\widetilde F=\{0\}$.

For the specified nonzero Pauli $q_v$ at each coordinate, choose $t_v$ with
$\omega(q_v,t_v)=1$.
The linear map $T_v$ taking the basis $q_v,t_v$ to $\ps Z,\ps X$ is invertible
and symplectic.
Set $T=\bigoplus_{v\in\{e\}\sqcup[n]}T_v$ and
\[
 \begin{aligned}
 \widehat W&=T(W),\\
 \mathcal Z&=T(\widetilde F)=\{(0,z):z\in\F^{n+1}\}.
 \end{aligned}
\]
Because the same invertible map acts on both $W$ and $\widetilde F$,
\[
 \begin{aligned}
 \widehat W\cap\mathcal Z
 &=T(W)\cap T(\widetilde F)\\
 &=T(W\cap\widetilde F)=\{0\}.
 \end{aligned}
\]

For the projection $\pi_x:\widehat W\to\F^{n+1}$ onto the $x$ component,
\[
 \begin{aligned}
 \ker\pi_x&=\widehat W\cap\mathcal Z=\{0\},\\
 \dim\widehat W&=\dim\F^{n+1}=n+1.
 \end{aligned}
\]
Thus $\pi_x$ is an isomorphism.
Writing $G$ for the matrix of the linear map $\pi_z\pi_x^{-1}$ gives
\[
 \widehat W=\{(a,Ga):a\in\F^{n+1}\}.
\]
Isotropy of $\widehat W$ implies, for every $a,b\in\F^{n+1}$,
\[
 \begin{aligned}
 0&=\omega((a,Ga),(b,Gb))\\
  &=a^{\mathsf T}(G+G^{\mathsf T})b,
 \end{aligned}
\]
and hence $G=G^{\mathsf T}$.
Since the adjacency matrix of a simple graph must have zero diagonal, we
remove the diagonal of $G$. Set $D_G=\operatorname{diag}(G_{vv})$ and apply
the coordinatewise map
\[
 C:(x,z)\longmapsto(x,z+D_Gx).
\]
Since $D_G=D_G^{\mathsf T}$, this map is symplectic, and it fixes every element
of $\mathcal Z$.
At each coordinate with $G_{vv}=1$, it simply applies the shear
$(x_v,z_v)\mapsto(x_v,z_v+x_v)$, which leaves the specified basis $\ps Z$
unchanged.
Consequently,
\[
 \begin{aligned}
 C(\widehat W)&=\{(a,\Gamma a):a\in\F^{n+1}\}=W_\Gamma,\\
 \Gamma&=G+D_G.
 \end{aligned}
\]
The matrix $\Gamma$ is symmetric with zero diagonal and hence is a graph
adjacency matrix.
The rank of the root projection is preserved by local invertible maps and
remains two, so the root is not isolated.
The root-zero part is also carried from $0\oplus S$ to $0\oplus S_\Gamma$.
The restriction of $CT$ to the physical coordinates is therefore the required
map, transforming both the code and the specified bases.
\end{proof}

Relabeling the strategy's bases and measurement outcomes by the same map
preserves its batches, number of rounds, and property of being perfect.
A necessary condition on the resulting graph can therefore be transferred
back to the original strategy.

\subsection{Completing Logical Fusion via Induced Paths}\label{sec:path}

For the upper bound, we use a procedure that finishes the remaining fusions
in at most one round after the first success.
An \emph{induced path} is a path whose vertices induce no edges other than the
path edges.
If such a path connects a successful coordinate to the root and its internal
vertices remain unfused, choosing failure basis $\ps X$ at those vertices
allows the success to yield recovery of the logical Bell parities.
Edges from the path to other vertices are allowed; the graph need not be a
tree.

Path-based measurement constructions appear in the tree-code strategy of
Rei{\ss} and van Loock~\cite[Sec.~V.F and Appendix~H.4]{reiss} and in the
publicly available distance-ordered construction for general connected graph codes~\cite{pr13260}.
The next lemma formulates and proves the related path-completion step in our
full outcome-tree model.

\Needspace{11\baselineskip}
\begin{lemma}[Completion along an induced path]\label{lem:path}
Consider the graph code of $\Gamma$.
Suppose that all previous failure bases were $\ps Z$, coordinate $i$ has
succeeded, and there is an induced path
\[
 P=(e=v_0,v_1,\ldots,v_\ell=i)
\]
of length $\ell$ (the number of edges), whose internal vertices have not yet
been fused.
Fuse all remaining coordinates in one round, using failure basis $\ps X$ at
the internal vertices of $P$ and $\ps Z$ elsewhere.
Logical fusion succeeds regardless of the outcomes of this last round.
If no coordinates remain, logical fusion has already succeeded.
\end{lemma}
Thus changing the bases inside the path suffices to complete logical fusion
without requiring another physical fusion to succeed.

\begin{proof}
Throughout this proof, let $N=S_\Gamma^{\perp_V}$.
Sum the graph-state generators $h_{v_j}$ separately over the even and odd
positions along the path:
\[
 \begin{aligned}
 w_{\mathrm{ev}}&=\sum_{\substack{0\le j\le\ell\\j\ \mathrm{even}}}h_{v_j},\\
 w_{\mathrm{odd}}&=\sum_{\substack{0\le j\le\ell\\j\ \mathrm{odd}}}h_{v_j}.
 \end{aligned}
\]
The two path neighbors of each internal vertex contribute to the same sum,
where their $\ps Z+\ps Z=0$ contributions cancel the internal $z$ component.
Because the path is induced, more distant path vertices contribute no
additional $\ps Z$ there.
The components along the path are therefore
\[
 \begin{array}{c|cc}
 \text{Coordinate}&w_{\mathrm{ev}}&w_{\mathrm{odd}}\\\hline
 e=v_0&\ps X&\ps Z\\
 v_j\ (0<j<\ell,\ j\text{ even})&\ps X&0\\
 v_j\ (0<j<\ell,\ j\text{ odd})&0&\ps X\\
 i=v_\ell\ (\ell\text{ even})&\ps X&\ps Z\\
 i=v_\ell\ (\ell\text{ odd})&\ps Z&\ps X
 \end{array}
\]
Outside the path, none of the selected generators contributes an $X$ component,
so only $0$ or $\ps Z$ can appear.
Each endpoint has only one path neighbor, so the cancellation at internal
vertices does not apply at the endpoints.

Let $L_{\mathrm{ev}}=w_{\mathrm{ev}}|_{[n]}$ and
$L_{\mathrm{odd}}=w_{\mathrm{odd}}|_{[n]}$ be the restrictions to the physical
coordinates.
For every $s\in S_\Gamma$, we have $(0,s)\in W_\Gamma$, so isotropy gives
\[
 \begin{aligned}
 \omega(L_{\mathrm{ev}},s)&=\omega(L_{\mathrm{odd}},s)=0,\\
 L_{\mathrm{ev}},L_{\mathrm{odd}}&\in N.
 \end{aligned}
\]
The two graph-state stabilizers commute, but their root components
anticommute:
\[
 \begin{aligned}
 0&=\omega_{\widetilde V}(w_{\mathrm{ev}},w_{\mathrm{odd}})\\
  &=\omega_e(\ps X,\ps Z)+\omega(L_{\mathrm{ev}},L_{\mathrm{odd}}),\\
 \omega(L_{\mathrm{ev}},L_{\mathrm{odd}})&=1.
 \end{aligned}
\]
Hence $[L_{\mathrm{ev}}],[L_{\mathrm{odd}}]$ form a basis of the
two-dimensional space $N/S_\Gamma$.
It is these logical Pauli classes that are independent;
$L_{\mathrm{ev}},L_{\mathrm{odd}}$ are their physical representatives.

Let $F_r$ be the final failure-basis span.
For $f\in F_r$, the component table above shows that $f$ commutes with both
$L_{\mathrm{ev}}$ and $L_{\mathrm{odd}}$ at every physical coordinate:
at internal path vertices, only $0$ and $\ps X$ occur; at the successful
coordinate $i$, we have $f_i=0$; and outside the path, only $0$ and $\ps Z$
occur. Hence
\[
 \omega(f,L_{\mathrm{ev}})=\omega(f,L_{\mathrm{odd}})=0.
\]
If also $f\in N$, its logical Pauli class is orthogonal to both elements of a
basis of $N/S_\Gamma$:
\[
 \begin{gathered}
 \bar\omega([f],[L_{\mathrm{ev}}])
 =\bar\omega([f],[L_{\mathrm{odd}}])=0\\
 \Longrightarrow\quad[f]=0.
 \end{gathered}
\]
Thus $F_r\cap N\subseteq S_\Gamma$, and logical fusion succeeds by
Lemma~\ref{lem:criterion}.
\end{proof}

For example, on the length-two path $e$--$a$--$i$, take
$w_{\mathrm{ev}}=h_e+h_i$ and $w_{\mathrm{odd}}=h_a$.
The two $\ps Z$ contributions at $a$ cancel, giving
$(w_{\mathrm{ev}})_a=0$ and $(w_{\mathrm{odd}})_a=\ps X$.
This explains the choice of failure basis $\ps X$ at the internal vertex.
For a length-one path, the interior is empty and the same proof applies.
Other coordinates may succeed in the same batch as $i$ or in the last round.
Each additional success merely sets the corresponding component of $F_r$
to zero and therefore preserves the orthogonality.
If no coordinates remain unfused, the same inclusion already holds for the
available $F_r$.

\section{Optimal Adaptive Fusion Rounds}\label{sec:schedules}

We characterize the batch sequences that can occur on an all-failure branch and use this characterization to prove the main theorem.
For necessity, we extract a graph from an arbitrary perfect strategy; for sufficiency, we construct a strategy using induced paths in that graph.
Together, these directions determine the minimum number of rounds.

\subsection{Characterization of Feasible Batch Sequences}

Partition $[n]$ into nonempty batches $T_1,\ldots,T_r$, and write
\[
  T_{>k}=T_{k+1}\cup\cdots\cup T_r.
\]
We ask when this sequence can occur on an all-failure branch of a perfect strategy.
The remaining batches may change after a success.

\begin{proposition}[Batch sequences on all-failure branches]\label{prop:schedules}
The following conditions are equivalent.
\begin{enumerate}
\item There exists a perfect strategy with an all-failure branch whose batches are $T_1,\ldots,T_r$, in that order.
\item There exists $\Gamma\in\GS$ such that, for every $k$ and every $i\in T_k$,
\begin{equation}\label{eq:later}
  \Nbr_\Gamma(i)\cap\bigl(\{e\}\cup T_{>k}\bigr)
       \ne\varnothing.
\end{equation}
\end{enumerate}
When these conditions hold, there exists a perfect strategy with the specified all-failure batch sequence that uses at most $r$ rounds on every branch.
\end{proposition}

Condition~\eqref{eq:later} requires each coordinate to be adjacent either to the root or to a coordinate fused in a later round.
The proposition states that this condition, in a suitable graph representation, is necessary and sufficient for a perfect strategy with the specified all-failure batch sequence.

\begin{proof}
\textbf{Necessity.}
Let $q$ be the failure-basis assignment along the specified all-failure branch.
Lemmas~\ref{lem:transverse} and~\ref{lem:graphform} put the code in graph form while mapping every basis in $q$ to $\ps Z$.
Suppose that condition~\eqref{eq:later} fails for some $i\in T_k$.
Then $i$ is not adjacent to $e$, so the restriction $s=h_i|_{[n]}$ belongs to $S_\Gamma$.

Follow the same history, including the measured values, until just before batch $T_k$.
Consider a branch on which only $i$ succeeds in that batch and every subsequent fusion fails.

Since condition~\eqref{eq:later} fails, all neighbors of $i$ belong to
$T_1\cup\cdots\cup T_k$ and have therefore been fused by the end of round $k$.
Their failure bases are all $\ps Z$, since the preceding history is unchanged
and the bases within $T_k$ were fixed before its outcomes were known.

Let $r'$ be the final failure-basis assignment on this branch.
Then
\[
  s_i=\ps X,\qquad s-\iota_i(\ps X)\in F_{r'}.
\]
By Lemma~\ref{lem:isolated}, logical fusion fails despite the physical success
at $i$, contradicting perfection.

\textbf{Sufficiency.}
In graph coordinates satisfying~\eqref{eq:later}, fuse the batches $T_1,T_2,\ldots$ in failure basis $\ps Z$ until the first round containing a success.
Let this be round $k$, and choose one successful coordinate $i\in T_k$.

Within the subgraph induced by $\{e,i\}\cup T_{>k}$, start at $i$ and repeatedly move to the root or to a neighbor in a later batch.
The batch index strictly increases at each step that does not reach the root, so condition~\eqref{eq:later} ensures that this process reaches $e$.
Choose a shortest path from $e$ to $i$ in this induced subgraph.
It is also an induced path in $\Gamma$, and all its internal vertices lie in $T_{>k}$ and have not yet been fused.
Lemma~\ref{lem:path} completes fusion in at most one additional round.
If $k<r$, the total is at most $k+1\le r$ rounds.
If $k=r$, condition~\eqref{eq:later} makes $i$ adjacent to $e$, and no coordinates remain.
The same lemma then gives success in $r$ rounds without an additional round.
The all-failure branch follows the prescribed $r$ batches.
Undoing the maps on the physical coordinates gives a perfect strategy for $S$.
\end{proof}

\subsection{Proof of the Main Theorem}

We use Proposition~\ref{prop:schedules} to prove the two inequalities in Theorem~\ref{thm:main}.
A graph extracted from an arbitrary perfect strategy gives a lower bound on its number of rounds.
Conversely, the distance layers of any graph representation give a strategy and hence an upper bound.
We then exclude perfect strategies for reducible codes.

\begin{proof}[Proof of Theorem~\ref{thm:main}]
\textbf{Lower bound.}
Take a perfect strategy using at most $R$ rounds, and choose an all-failure branch with $r\le R$ batches.
Proposition~\ref{prop:schedules} gives a graph $\Gamma\in\GS$ satisfying~\eqref{eq:later}.
From any $i\in T_k$, following neighbors in later batches reaches $e$ in at most $r-k+1$ edges.
Thus
\[
  \min_{\Gamma'\in\GS}\ecc_{\Gamma'}(e)
       \le \ecc_\Gamma(e)\le r\le R.
\]
Minimizing the worst-case round count $R$ over perfect strategies proves the lower bound.

\textbf{Upper bound.}
If $S$ is irreducible, Lemma~\ref{lem:connected} implies that every graph in $\GS$ is connected.
Choose one with root eccentricity $h$.
For $1\le k\le h$, define the batches from the outermost distance layer inward:
\begin{equation}\label{eq:layers}
  T_k=\{i:\dist_\Gamma(e,i)=h-k+1\}.
\end{equation}
These batches partition $[n]$.
A shortest path to a vertex at distance $h$ visits every distance $1,\ldots,h$, so no batch is empty.
Every vertex at distance at least two has a neighbor one step closer to the root, in a later batch.
Every vertex at distance one is adjacent to $e$.
Hence condition~\eqref{eq:later} holds, and Proposition~\ref{prop:schedules} gives a perfect strategy using at most $h$ rounds.
Minimizing $h$ over $\GS$ gives
\[
  R^*(S)\le\min_{\Gamma\in\GS}\ecc_\Gamma(e).
\]
Together with the lower bound, this yields equality in~\eqref{eq:main}.

\textbf{Reducible codes.}
Suppose that a reducible $S$ admitted a perfect strategy.
The graph extracted in the lower-bound argument would satisfy~\eqref{eq:later}.
Following neighbors in later batches would connect every physical vertex to $e$, making the graph connected.
This contradicts Lemma~\ref{lem:connected}, since coordinatewise maps preserve irreducibility.
\end{proof}

The distance-layer construction also has a concrete physical interpretation.
Consider the four-outcome measurement of a photon pair with rank-one
projectors onto $|00\rangle$, $|11\rangle$, $|\Psi^+\rangle$, and
$|\Psi^-\rangle$, where
$|\Psi^\pm\rangle=(|01\rangle\pm|10\rangle)/\sqrt{2}$.
The first two outcomes are fusion failures, reporting equal individual
$\ps Z$ values on both photons; the latter two are successes, reporting
two Bell parities. Applying the same single-qubit Clifford change of basis
to both photons gives the corresponding measurement for any chosen Pauli
failure basis. Thus the physical outcomes supply the Pauli information
assumed in Section~\ref{sec:roundmodel}.

\Needspace{8\baselineskip}
Recovering both logical Bell parities is equivalent to unambiguously
distinguishing the four encoded Bell states.
To evaluate the logical success probability, choose each of the four
encoded Bell inputs
$(|\bar 0\bar 0\rangle\pm|\bar 1\bar 1\rangle)/\sqrt{2}$ and
$(|\bar 0\bar 1\rangle\pm|\bar 1\bar 0\rangle)/\sqrt{2}$
with equal probability, where $|\bar 0\rangle,|\bar 1\rangle$ are an
orthonormal logical basis of the code. An outcome branch identifies its
input unambiguously if its outcomes have positive Born probability for
exactly one of these four Bell states. Their uniform mixture is
$\rho_S\otimes\rho_S$, with $\rho_S=\Pi_S/2$ and $\Pi_S$ the projector
onto the code space. In graph coordinates, every failure basis along the
all-failure branch of the distance-layer strategy is $\ps Z$.
A graph code has no nonidentity purely $\ps Z$-type stabilizer, so the
complete $\ps Z$-outcome string $x\in\{+1,-1\}^n$ of each block is
uniformly distributed, with $p(x)=2^{-n}$. The two blocks are independent
under $\rho_S\otimes\rho_S$, and all fusions fail precisely when their
$\ps Z$-outcome strings agree. Consequently,
\[
 \Pr(\mathrm{all\ fail})=\sum_x p(x)^2=2^{-n}.
\]

The distance-layer strategy recovers both logical Bell parities on every
branch containing a physical success, by the perfection already proved
above. On an all-failure branch with failure-basis assignment $q$,
Lemma~\ref{lem:full} gives a nontrivial logical Pauli
$w\in(F_q\cap N)\setminus S$. The corresponding Pauli operator on the
first block commutes with every measured projector, while mapping each
encoded Bell state to a different one. Thus the branch has the same
probability for two distinct Bell inputs and cannot identify the logical
Bell state. Hence, for the four encoded Bell inputs chosen uniformly,
the average probability of unambiguously identifying the logical Bell
state is exactly
$1-2^{-n}$. Thus the round-optimal distance-layer strategy of Theorem~\ref{thm:main}
is physically realizable with this fusion measurement and attains this
success probability. The theorem's round-count lower bound concerns
perfect strategies in the full outcome-tree model of
Section~\ref{sec:roundmodel}.

\section{Examples}\label{sec:example}

Two codes illustrate what the main theorem optimizes.
The first shows how changing the graph representation reduces the number of rounds in a strategy built from distance layers.
The second requires at least two rounds for any perfect strategy in our full outcome-tree model.

\subsection{Optimization over Graph Representations}

Consider the path $e$--$1$--$2$--$3$.
Its root eccentricity is three, giving the distance-layer batches $\{3\},\{2\},\{1\}$.
For example, if coordinate $3$ succeeds first, fuse the remaining coordinates $1,2$ in one round with failure basis $\ps X$.
If the first two rounds fail and coordinate $1$ succeeds in the third, the success at a neighbor of the root completes logical fusion.

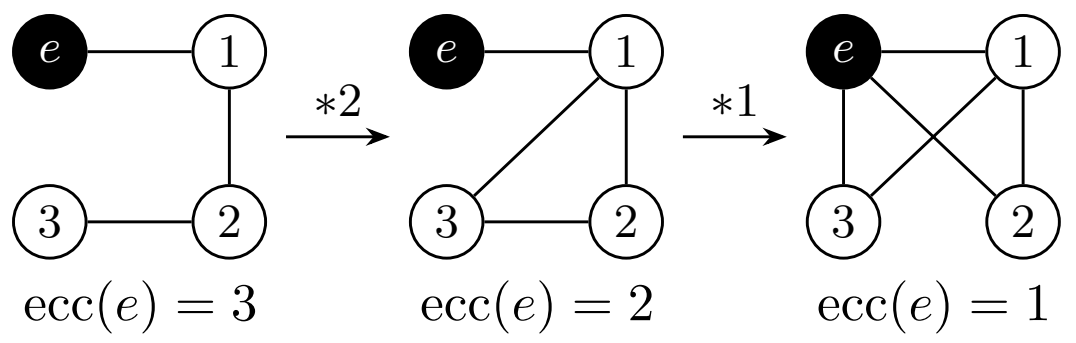
\begin{figure}[htbp]
\centering
\begin{tikzpicture}[
  scale=1.25,
  v/.style={circle,draw,fill=white,inner sep=0pt,minimum size=4.8mm,font=\footnotesize},
  root/.style={v,fill=black,text=white},
  every path/.style={line width=0.55pt}
]
  \begin{scope}
    \node[root] (a0) at (0,0.9) {$e$};
    \node[v] (a1) at (0.95,0.9) {$1$};
    \node[v] (a2) at (0.95,0) {$2$};
    \node[v] (a3) at (0,0) {$3$};
    \draw (a0)--(a1)--(a2)--(a3);
    \node[font=\small] at (0.48,-0.45) {$\ecc(e)=3$};
  \end{scope}
  \draw[-{Stealth[length=1.6mm]}] (1.25,0.45)--(1.8,0.45)
    node[midway,above,font=\footnotesize] {$*2$};
  \begin{scope}[xshift=2.1cm]
    \node[root] (b0) at (0,0.9) {$e$};
    \node[v] (b1) at (0.95,0.9) {$1$};
    \node[v] (b2) at (0.95,0) {$2$};
    \node[v] (b3) at (0,0) {$3$};
    \draw (b0)--(b1)--(b2)--(b3)--(b1);
    \node[font=\small] at (0.48,-0.45) {$\ecc(e)=2$};
  \end{scope}
  \draw[-{Stealth[length=1.6mm]}] (3.35,0.45)--(3.9,0.45)
    node[midway,above,font=\footnotesize] {$*1$};
  \begin{scope}[xshift=4.2cm]
    \node[root] (c0) at (0,0.9) {$e$};
    \node[v] (c1) at (0.95,0.9) {$1$};
    \node[v] (c2) at (0.95,0) {$2$};
    \node[v] (c3) at (0,0) {$3$};
    \draw (c0)--(c1)--(c2)--(c0)--(c3)--(c1);
    \node[font=\small] at (0.48,-0.45) {$\ecc(e)=1$};
  \end{scope}
\end{tikzpicture}
\caption{Three graph representations of the same code.
The operation $*v$ denotes local complementation at vertex $v$.
The root eccentricity decreases from three to two to one.
Although the original path has three nonempty distance layers, the theorem gives a perfect one-round strategy.}
\label{fig:path}
\end{figure}

This three-round construction is not optimal for the code.
Local complementation at vertex $2$ adds edge $13$, reducing the root eccentricity to two.
Local complementation at vertex $1$ then adds edges $e2,e3$ and removes edge $23$.
The root is now adjacent to all three physical vertices (Fig.~\ref{fig:path}).
This representation has root eccentricity one, so the same code has $R^*=1$.
The failure bases for the one-round strategy on the original code are obtained by undoing the coordinatewise maps.
Thus the strategy from the original graph uses three rounds in the worst case, whereas optimization over the graph orbit gives a single round without feed-forward.

\subsection{A Code Requiring Two Fusion Rounds}\label{sec:octahedron}

Consider the octahedral graph $K_{2,2,2}$ with parts $\{e,c\},\{a,b\},\{d,f\}$ and root $e$.
It is connected, so the corresponding five-qubit code is irreducible.
In physical-coordinate order $(c,a,b,d,f)$, its stabilizer subspace is
\[
 \begin{aligned}
 S=\spn\{&\ps{XZZZZ},\ps{IXXII},\\
         &\ps{IYZYZ},\ps{IYZZY}\}.
 \end{aligned}
\]
This code is locally Clifford equivalent to the five-qubit example in
Ref.~\cite[Section~5]{goodenough} that admits no perfect nonadaptive fusion strategy.
We use it to illustrate Theorem~\ref{thm:main} by determining the
minimum root eccentricity over its graph orbit.

\paragraph{Root eccentricity.}
This graph has $\ecc_\Gamma(e)=2$, since $c$ is the only physical vertex not adjacent to $e$.
To rule out eccentricity one, start from $K_{2,2,2}$ and repeatedly
apply local complementation at each of the six labeled vertices,
retaining each distinct graph until the orbit is closed.
The orbit contains $40$ labeled graphs; their root degrees are
$1,2,3,4$ with respective multiplicities $10,8,12,10$.
In particular, the root is never adjacent to all five physical vertices.
Thus $\min_{\Gamma'\in\GS}\ecc_{\Gamma'}(e)=2$, and
Theorem~\ref{thm:main} gives $R^*(S)=2$.

\paragraph{Explicit strategy.}
First fuse $c$ in failure basis $\ps Z$, and choose the second round according to its outcome.
\begin{itemize}
\item If $c$ fails, fuse $a,b,d,f$ in failure basis $\ps Z$.
All four vertices are adjacent to the root.
If any one succeeds, the length-one path from that coordinate to the root satisfies Lemma~\ref{lem:path}, so logical fusion succeeds.
\item If $c$ succeeds, use the induced path $e$--$a$--$c$ and fuse $a$ in failure basis $\ps X$ and $b,d,f$ in failure basis $\ps Z$.
Lemma~\ref{lem:path} gives recovery for every combination of successes and failures in the second round.
\end{itemize}
This strategy fuses all coordinates in two rounds on every branch, and logical fusion succeeds whenever at least one physical fusion succeeds.
It realizes the optimum $R^*=2$ given by Theorem~\ref{thm:main}; one feed-forward step between fusion rounds is necessary in our model.

\section{Conclusion}\label{sec:conclusion}

For an irreducible $[[n,1]]$ stabilizer code, we have shown that the minimum
worst-case number of measurement rounds for perfect fusion equals the minimum
root eccentricity over its rooted graph representations. The lower bound applies
to arbitrary perfect strategies that adapt their batches and failure bases to
previous outcomes. We also give a necessary and sufficient graph adjacency
condition for a batch sequence to occur on an all-failure branch of a perfect
strategy.

These results determine the optimal number of measurement rounds for fusion
strategies that allow FF and provide both a strategy attaining that number and
a proof that fewer rounds cannot be perfect. They do not directly optimize
hardware latency: measurement times, FF processing times, and storage costs
between rounds are not specified.

The results apply to the full outcome-tree model of Section~\ref{sec:roundmodel},
which includes zero-probability branches. Restricting to
positive-probability branches for a particular input state and measurement
device gives a different optimization problem.

\section*{Acknowledgments}
This work was supported in part by JST Moonshot R\&D Grant Number JPMJMS256F
and in part by the MITOU Target Program of the Information-technology
Promotion Agency, Japan (IPA).

\section*{Author Contributions}
Seita Tsuboi led the research. Takuya Kasamura, Leo Itoh, Motoharu Tsuboi,
and Tomohiro Hatori contributed to research discussions and to reviewing
and verifying the mathematical arguments.
Junichiro Kadomoto supervised the research.

Large language models were used for mathematical exploration and checking,
assistance with code and calculations, literature checking, drafting and revising
the manuscript, and preparing LaTeX and figures. The tools used include Claude
and OpenAI Codex.

\end{document}